\documentclass[aps,prl,twocolumn,groupedaddress,floatfix,showpacs]{revtex4}

\usepackage{graphicx}

\begin{document}

\title{Angular momentum of a strongly focussed Gaussian beam}

\author{Timo A. Nieminen}
\email[]{timo@physics.uq.edu.au}

\author{Norman R. Heckenberg}
\author{Halina Rubinsztein-Dunlop}

\affiliation{Centre for Biophotonics and Laser Science, Department of Physics,
The University of Queensland, Brisbane QLD 4072, Australia}

\begin{abstract}
A circularly polarized rotationally symmetric paraxial laser beams carries
$\pm\hbar$ angular momentum per photon as spin. Focussing the beam with a
rotationally symmetric lens cannot change this angular momentum flux, yet
the focussed beam must have spin $|S_z| < \hbar$ per photon. The remainder
of the original spin is converted to orbital angular
momentum, manifesting itself as a longitudinal optical vortex at the focus.
This demonstrates that optical orbital angular momentum can be generated
by a rotationally symmetric optical system which preserves the total angular
momentum of the beam.
\end{abstract}

\maketitle 

\section{Introduction}

That optical and other electromagnetic fields can carry angular momentum is
a direct result of the fact that they can carry linear momentum. Since the
linear momentum flux density of an electromagnetic field is
$\mathbf{E}\times\mathbf{H}/c$, giving
\begin{equation}
\mathbf{p} = \frac{1}{2c} \mathrm{Re}(\mathbf{E}\times\mathbf{H}^\ast)
\end{equation}
for the time-averaged momentum flux density of a time-harmonic electromagnetic
beam, it is tempting to write
\begin{equation}
\mathbf{j} = \frac{1}{2c}
\mathrm{Re}(\mathbf{r}\times(\mathbf{E}\times\mathbf{H}^\ast)).
\end{equation}
as the corresponding angular momentum flux density. However, since the
conservation law for linear momentum only contains the time-derivative of
the momentum density, and the divergence of the flux (or the volume integrals
of these, noting that the volume integral of a divergence of a quantity
can be written as the surface integral of the quantity), it is premature to
identify this as the actual angular momentum flux density.
If the above expression was in fact the correct angular momentum flux
density, then the angular momentum of a circularly polarized plane would be
zero. Since the correct classical angular momentum density must agree with
the classical limit of the quantum angular momentum density, this must be
incorrect. This was recognized long ago, and a separation of
the angular momentum of the field can be separated into
spin and orbital components yields a result in agreement with the
quantum result~\cite{humblet1943,vanenk1994,crichton2000}.

The total momentum or angular momentum flux through a plane can be
unambiguously determined
by integrating the above quantities over the plane; this is a straightforward
application of the integral form of the conservation law for angular momentum,
and is therefore independent of the actual choice of expression for the
angular momentum density. It can be noted that the
actual momentum and angular momentum densities (as opposed to the flux
densities given above) contain an extra $1/c$ factor, and can be integrated
over the whole beam to give the total momentum and angular momentum content.
However, for the case of an infinitely long beam,  these quantities are
infinite, and it is better to consider only the fluxes.

Separating the total angular momentum flux of the beam into spin and orbital
components gives
\begin{equation}
\mathbf{J} = \mathbf{L} + \mathbf{S}
\end{equation}
where the total angular momentum flux
$\mathbf{J}$ is the sum of an orbital term $\mathbf{L}$ and
a spin term $\mathbf{S}$. A similar relationship exists for the angular
momentum flux densities and densities at any point. The distinction is simply
that the spin angular momentum density is independent of the choice of origin;
that is, it is invariant with respect to translations of the coordinate
system. Since the local conservation of angular momentum as the beam
propagates in free space cannot depend on the choice of the origin of the
coordinate system, both the orbital and spin components must be individually
conserved, as well as the total angular momentum, since there would otherwise
exist a choice-of-origin dependent torque exerted by the beam on free space.
These quantities can only
change if the field interacts with matter~\cite{vanenk1994}.

In general, the separation into spin and orbital components is not quite as
straightforward as one might hope~\cite{humblet1943,vanenk1994,crichton2000}.
One can, however, write expressions for the spin and orbital components.
The Cartesian components of the time-averaged
spin angular momentum flux density $\mathbf{s}$
are~\cite{humblet1943,vanenk1994,crichton2000}
\begin{eqnarray}
s_x & = & \epsilon_0 \mathrm{Im}( E_y E_z^\star ) / \omega \nonumber \\
s_y & = & \epsilon_0 \mathrm{Im}( E_x^\star E_z ) / \omega \\
s_z & = & \epsilon_0 \mathrm{Im}( E_x E_y^\star ) / \omega \nonumber
\end{eqnarray}
where $E_{x,y,z}$ are the Cartesian components of $\mathbf{E}$, the
complex vector amplitude. This result can also be written in terms of the
Levi--Civita symbol as $s_i = \mathrm{i} \epsilon_0 \epsilon_{ijk} E_j
E_k^\star / (2\omega)$, where the expression is summed over repeated indices
and the real part is taken. The orbital components
are~\cite{humblet1943,vanenk1994,crichton2000}
\begin{equation}
l_i = \mathrm{i} \epsilon_0 E_j (\mathbf{r}\times\nabla) E_j^\star
/ (2\omega).
\end{equation}
These give simple results for paraxial beams. For
example, a circularly polarized paraxial beam of power $P$ and frequency
$\omega$ has a spin angular momentum flux of $\pm P/\omega$, depending
on the handedness of the polarization~\cite{jackson1999book}.
This result is equivalent to the
quantum mechanical result of $\pm \hbar$ per photon. If the beam has a uniform
phase over a plane, or, more generally, phase that is rotationally symmetric
about the beam axis, then the orbital angular momentum flux density is zero.

We will consider such rotationally symmetric beams here, and present specific
results for a Gaussian beam.

\section{Angular momentum of a finite beam}

The case of spin angular momentum flux equal to $P/\omega$ results only
in the paraxial approximation, as it depends on $E_z$ being zero. If we
consider a beam of finite width in its focal plane, then the beam will
spread through diffraction, and will, at a sufficiently large distance,
be propagating in a purely radial direction. That is, for large $r$,
we must have $E_r = 0$.
In this case, the electric field is purely tangential, and the spin
angular momentum density in polar spherical coordinates is 
\begin{equation}
s_r = \epsilon_0 \mathrm{Im}( E_\theta E_\phi^\star ) / \omega,
\end{equation}
with the other vector components being zero. For a rotationally symmetric
beam of the type we consider here, $s_r$ will be independent of the
azimuthal angle $\phi$.

Therefore, the maximum possible contribution to the total spin angular
momentum, of which, by symmetry, only the $z$ component is non-zero,
is $s_r \cos\theta$, where $\theta$ is the angle measured from the $z$ axis.
Integrating this over the beam must result in $|S_z| < \hbar$ per photon.
A consequence of this is that a focussed beam cannot be purely circularly
polarized everywhere in a plane, including the focal plane.
If the beam is rotationally symmetric and locally circularly
polarized in the far field, the beam cannot be partially plane polarized in
the focal plane, either---therefore, the electric field must have an
axial component ($E_z \ne 0$). A similar argument using linear momentum shows
that plane polarized beams must have $E_z \ne 0$ in the focal plane as well.

\begin{figure}[htb]
\centerline{\includegraphics[width=\columnwidth]{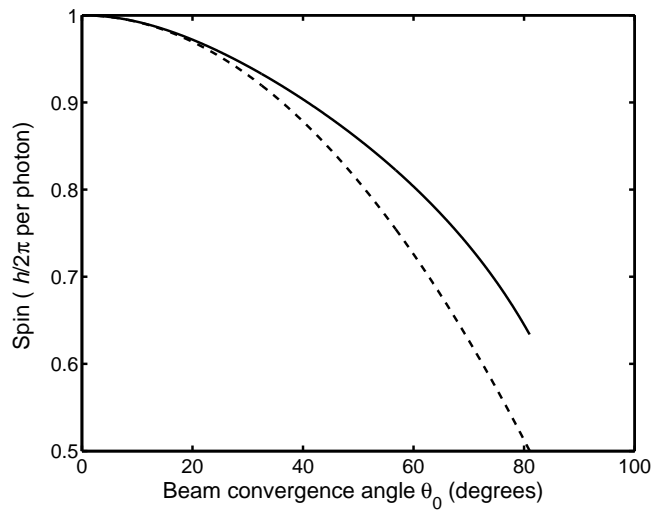}}
\caption{Spin angular momentum in $\hbar$ per photon for a non-paraxial
TEM$_{00}$ Gaussian beam. The solid line shows the exact result, the
dotted line shows the small-angle approximation.}
\label{spin_fig}
\end{figure}

As an example, we will find the spin carried by a focussed Gaussian beam.
A paraxial Gaussian beam has a (scalar) amplitude in the far field of
\begin{equation}
U = U_0 \exp\left[ - \frac{k^2 w_0^2 \tan^2\theta}{4} \right]
\end{equation}
where $k$ is the wavenumber and $w_0$ is the paraxial beam
waist radius~\cite{nieminen2003a}, or in terms of the beam convergence angle
given by $\tan\theta_0 = 2/(kw_0)$,
\begin{equation}
U = U_0 \exp( - \tan^2\theta / \tan^2\theta_0 ).
\end{equation}
For maximum possible spin, we have $s_r = \epsilon_0 U^2 /\omega$, and
the total spin angular momentum of the beam,
in units of $\hbar$ per photon, can be found by integrating over a hemisphere:
\begin{equation}
S_z = A/P
\end{equation}
where
\begin{equation}
A = \int_0^{\pi/2} \exp( - 2 \tan^2\theta / \tan^2\theta_0 ) \sin\theta
\cos\theta \mathrm{d}\theta
\end{equation}
and
\begin{equation}
P = \int_0^{\pi/2} \exp( - 2 \tan^2\theta / \tan^2\theta_0 ) \sin\theta
\mathrm{d}\theta.
\end{equation}
This can be readily integrated numerically, and the result for
all practically realizable Gaussian beams is shown in
figure~\ref{spin_fig}. The qualitative behaviour seen here is expected
since $S_z = 0.5 P/\omega$ for a dipole radiation
field~\cite{crichton2000}.

For $\theta_0 \ll 1$, small angle approximations can be made, giving
\begin{equation}
S_z = 1 - \theta_0^2/4.
\end{equation}
The relative error in the change in spin ($1 - S_z$) is less than $0.01$
for beam convergence angles $\theta_0 < 6.25^\circ$, and less than $0.1$
for $\theta_0 < 21^\circ$.

Although the spin angular momentum flux of the beam is reduced by the beam
being focussed, a lossless rotationally symmetric optical system cannot
change the total angular momentum flux of an electromagnetic
field~\cite{waterman1971}. Therefore, there must be a corresponding increase
in the orbital angular momentum. This is in remarkable contrast to the
usual methods of generating optical vortices which employ astigmatic
or cylindrical lenses or holograms designed to break the rotational symmetry.
The key difference is that orbital angular momentum generation by focussing
depends on the initial presence of spin angular momentum, whereas astigmatic
systems do not.

An interesting question that remains to be answered is in what way the
focussed beam carries the orbital angular momentum. This is best addressed
by considering a rigorous electromagnetic model of the beam.

\subsection{Multipole expansion}

A time-harmonic electromagnetic beam can be represented as a sum of
of electric and magnetic multipoles:
\begin{equation}
\mathbf{E}(\mathbf{r}) = \sum_n^\infty \sum_{m=-n}^n
a_{nm} \mathbf{M}_{nm} + b_{nm} \mathbf{N}_{nm}
\end{equation}
where $\mathbf{M}_{nm}$ and $\mathbf{M}_{nm}$ are the TE and TM
regular multipole fields, or vector spherical
wavefunctions~\cite{nieminen2003a}. Not only are these wavefunctions a
complete orthogonal set of divergence-free solutions of the vector
Helmholtz equation (and hence solutions to the Maxwell equations),
they are also eigenfunctions of the angular momentum
operator $J^2$, with eigenvalues $[n(n+1)]^{1/2}$, and $J_z$, with
eigenvalues $m$. The spin and orbital contributions to the anglular
momentum can be calculated from the expansion coefficients $a_{nm}$
and $b_{nm}$~\cite{crichton2000,bishop2003}.

The only non-zero multipole coefficients for a left-circularly polarized
rotationally symmetric beam are those with $m=1$. Thus, the total angular
momentum about the $z$ axis is $\hbar$ per photon. The multipole expansion
coefficients for the beam can be determined by an overdetermined
point-matching method~\cite{nieminen2003a}. For a beam of finite
width, it is found that the total spin is less than $\hbar$ per photon
(spin calculated in this way exactly reproduces the curve in
figure~\ref{spin_fig}). The remainder of the angular momentum is orbital.

\begin{figure}[htb]
\centerline{\includegraphics[width=\columnwidth]{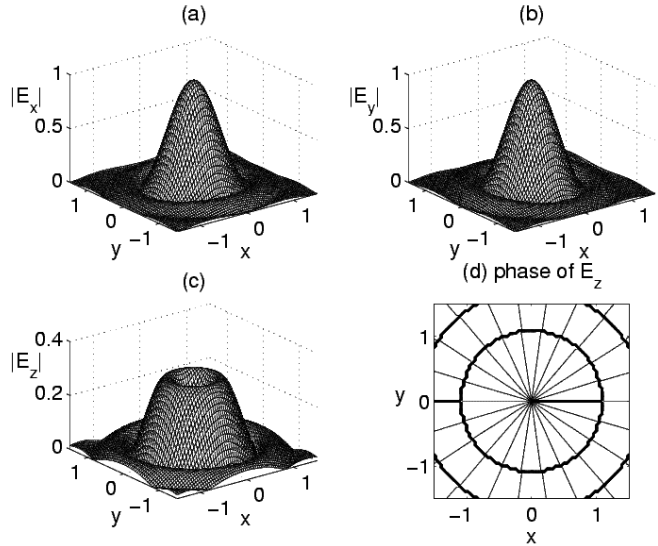}}
\caption{Electric field components of a strongly focussed circularly
polarized Gaussian beam, with a convergence angle of 45$^\circ$. The
$x$, $y$, and $z$ components of the electric field in the focal plane
are shown in (a), (b), and (c), while (d)
shows phase contours (with a spacing of $2\pi/20$)
for the $z$ component, showing azimuthal variation
of phase as seen in vortex beams.}
\label{fields_fig}
\end{figure}

Since the multipole expansion of the beam is known, the fields can be
calculated at any point in space. The components of the electric field in
the focal plane are shown in figure~\ref{fields_fig}. For a strongly focussed
beam such as is shown in figure~\ref{fields_fig}, the longitudinal (ie $z$)
component of the field is significant, with a magnitude of $\approx 0.3$
times the transverse components. All components of the electric field
show secondary diffraction rings (note that the radial dependence of
multipole fields includes a spherical Bessel function). The phases of the
$x$ and $y$ components are uniform, except for an increment of $\pi$ between
successive diffraction rings, and, due to the circular polarization, differ
by $\pi/2$ from each other. The phase of the $z$ component, however, shows
a clear azimuthal dependence identical to that seen in $l = 1$
paraxial vortex modes. Since this vortex behaviour is only possessed by the
longitudinal component of the field, this can be called a \emph{longitudinal
optical vortex}.

Calculation of the Poynting vector shows that there is indeed a transverse
component, which, since its handedness is uniform, is responsible for the
transport of the orbital angular momentum.

\begin{figure}[htb]
\centerline{\includegraphics[width=0.8\columnwidth]{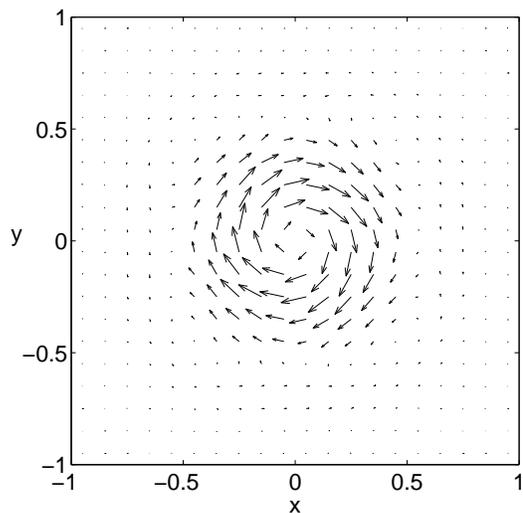}}
\caption{Poynting vector of a strongly focussed circularly
polarized Gaussian beam, with a convergence angle of 45$^\circ$. The
transverse part ($x$ and $y$ components) of the Poynting vector in the
focal plane are shown.}
\label{vortex_fig}
\end{figure}

As the beam is more strongly focussed, the magnitude of the longitudinal
($z$) component of the field increases, and the orbital angular momentum
increases as a result. The same increase can also be considered to result
from the decrease of spin angular momentum, along with the conservation
of total angular momentum. The change in the angular momentum and the
growth of the longitudinal optical vortex is smooth and well-behaved as
the convergence angle of the beam is increased, with
no sudden qualitative or quantitative changes. As the beam is more strongly
focussed, the diffraction rings also become more prominent, but this does
not affect the angular momentum of the beam.

\section{Discussion}

%
%
\subsection{Torque density acting on lens}

In light of the above considerations, the transformation of spin angular
momentum to orbital angular momentum by the lens must result in a
reaction torque density about the $z$-axis acting on the lens that depends
on the choice of origin. While such a torque acting on empty space is
unacceptable on physical grounds, it is not only entirely reasonable, but
expected, in the case of a lens.

One need only to consider a simple ray picture of the action of a lens,
not centered on the $z$ axis, on a ray parallel to the $z$-axis. The
focussed ray will generally not pass through the $z$-axis, and is not
parallel to the $z$-axis, and hence carries orbital angular momentum
about the $z$-axis. Consequently, there must be a reaction torque density
acting on the lens. Note that this torque density depends on the choice
of origin. The component about any axis parallel to the beam axis of the
total torque acting on the lens is, of course, zero.

\subsection{Rotation in optical traps}

The presence of orbital angular momentum in the focal region suggests that
orbital motion of absorbing or reflective spherical particles should be
observable in optical tweezers. However, since particles trapped
in a Gaussian beam trap will be located on the beam axis, all that
will be observed will be spinning of the particle about its axis. In order to
observe this orbital angular momentum, it would be necessary to use a
multi-ringed beam with zero orbital angular momentum,
for example a Laguerre-Gauss mode LG$_{p0}$, with $p>1$, or a Bessel beam,
as the input to the objective lens of the trap. In this case, particles
trapped in one of the rings rather than the central spot can be be expected
to undergo orbital motion.

%

\section{Conclusion}

We note that focusing a circularly polarized beam preserves the total angular
momentum flux of the beam about its axis. However, the spin component of the
angular momentum flux is necessarily reduced as the beam is more strongly
focussed. Due to the conservation of total angular momentum when the beam is
focussed by a rotationally symmetric optical system, there must be a
corresponding increase in the orbital angular momentum flux.
This result is remarkable in that it predicts the
generation of orbital angular momentum by a rotationally symmetric optical
system, in apparent contradiction with common expectation.

This orbital
angular momentum is carried by the axial component of the electric field,
$E_z$, which has the typical $\exp(\mathrm{i}\phi)$ dependence of charge 1
optical vortices; we call this a longitudinal optical vortex.


\end{document}